\documentclass[preprint,showpacs]{revtex4}
\usepackage{epsf,graphicx}
\begin{document}

\title{Universal Scaling Function of Velocity Rotation in Spiral galaxies}

\author{Reza Torabi}
\affiliation{Physics Department, Tafresh University, P.O.Box:
39518-79611, Tafresh, Iran} \email{rezatorabi@aut.ac.ir}

\begin{abstract}
The data of velocity rotation curve in spiral galaxies, almost for
galaxies which have close central surface brightness, collapse onto
a universal scaling function. Since scaling functions are the
signature of emergence in complex systems we are led to the idea
that explanation of constant velocity in spiral galaxies needs
considering cooperative behavior instead of interpretation based on
reductionism approach.
\end{abstract}


\maketitle

Velocity rotation curve of spiral galaxies, hereafter RC, is one of
the wonderful and important subjects in physics. The curve
represents the rotation velocity of stars in a galaxy versus their
distance, $r$, from the galaxy center. The rotation velocity of
spiral galaxies increases with $r$ and tends to a constant value.
RCs do not show any Keplerian fall-off \cite{Salpeter, Sancisi,
Krumm, Moorsel, Rohlfs} for large radii. This behavior forced
physicists to present new theories and interpretations. The first
interpretation is that the luminous disk is embedded within a halo
of dark matter \cite{Zwicky, Rubin, Bosma, Ashman}. The distribution
of dark matter in galaxies should be proportional to $r$ in large
radii regime to have a constant velocity. In spite of intensive
search for the components of non-baryonic dark matter, no candidate
particles have been observed. Scientists have tried to search for
modified gravity theories to explain galaxy dynamics without the
need for exotic dark matter as the second interpretation. These
theories are based on the modification of Newtonian dynamics
\cite{Milgrom, Sanders} or generalization of Einstein's general
relatively \cite{Brownstein, Moffat}. Study of galaxies and their
RCs helps us with better understanding of the universe and its
component matter or of the unknown gravitational force. In the other
words, galaxies are the big labs of the universe to study gravity
and matter.

Most of the empirical relations of galaxies are scaling laws. One of
the most firmly established empirical scaling relations of disk
galaxies is the Tully-Fisher relation \cite{Tully, Schneider}. This
relation indicates a tight correlation between the total luminosity
and the maximum rotation velocity of the galaxy, however a complete
physical interpretation of this relation is still missing
\cite{Courteau}. Tully-Fisher relation shows a notable dependence on
morphological types \cite{Shen} and distance estimation  is one of
its applications \cite{Tully}. In this letter other scaling
relations are established beside conventional scaling laws and it is
shown that the data of RCs, almost for galaxies which have close
central surface brightness, collapse onto a universal scaling
function. This behavior can help us to obtain a better understanding
of galaxy formation and dynamics. Also, the universal function
enables us to estimate an unknown astrophysical parameter by knowing
other parameters.

Tully-Fisher relation has less dispersion in I-band luminosity and
also we are mainly interested in mass rather than luminosity, so we
concern ourselves with the Tully-Fisher and other scaling relations
measured in this band \cite{Courteau}. In this letter the distance
and RCs data have been derived from Persic et al. paper
\cite{Persic}. Furthermore, the luminosity and central surface
brightness in I-band are read from a database which is related to
the Courteau et al. paper \cite{Courteau}. Note that the Hubble
constant is assumed $H_0=70\; km\; s^{-1}$ entire this letter.

In order to find scaling relations logarithmic scale is used for
diagrams. RCs in logarithmic scale are linear for small radii
(linear regime) while the velocity reaches to a constant value for
large radii (constant regime). We draw RCs for different galaxies
and then pick a bunch of them that have almost the same linear
regimes (Fig. 1). It is seen that the galaxies of each selected
bunch have close I-band central surface brightness, $\mu_{0I}$ (e.g.
see table. 1). This indicator (close central surface brightness)
works well specially for later types of spiral galaxies. To have a
view about other bunches we can have a comparison with the selected
bunch in Fig. 1. RCs related to bunches with smaller central surface
brightness are placed on the right hand of the RCs in Fig. 1 while
those with larger central surface brightness will be on its left
hand. It will be shown that the rotation curve data of different
bunches of galaxies show a universal scaling behavior with different
scaling exponents.
\begin{figure}
\centering
\includegraphics[scale=0.5]{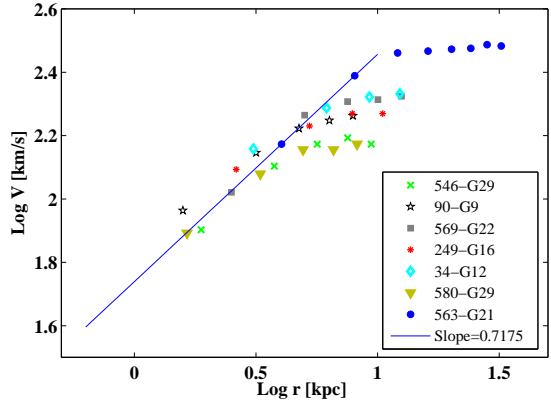}
\caption{RCs for a bunch of galaxies which have close central
surface brightness.}
\end{figure}

Linear regime indicates a power low behavior for rotation velocity
in small radii as
\begin{equation}
V\sim r^\beta \;\;\;\;\; r\ll r_x,
\end{equation}
where the slope is $\beta=0.7175$ for the selected galaxies of
table. 1. Bunches of galaxies with larger central surface brightness
have a little larger $\beta$ and vise versa. Also, this exponent
shows morphological dependence. Bunches of earlier types of spiral
galaxies have a  minute larger slope in linear regime. $r_x$ is the
radius in which the velocity reaches to its constant value. We call
this radius, the crossover radius and is obtained by intersecting
the fitted lines for two regimes, linear and constant. The method
for determination of $r_x$ is similar to the one depicted by Family
and Vicsek to define the crossover time in surface growth
\cite{Barabasi}.

\begin{table}
\caption{Characteristics of selected bunch of galaxies}
\label{table:tab1}
\begin{tabular}{ c c c c c c } \hline
\toprule
$Galaxy\;\;\;\;$ & $Morph\;\;\;\;$ & $\mu_{0I} [\frac{mag}{arcsec^2}]\;\;\;\;$ & $\log L_I [L_{\odot}]$ \\
\hline
$546-G29\;\;\;\;$ & $Sc\;\;\;\;$ & $18.82\;\;\;\;$ & $10.231\;\;\;\;$ \\
\hline
$563-G21\;\;\;\;$ & $Sbc\;\;\;\;$ & $18.73\;\;\;\;$ & $11.241\;\;\;\;$ \\
\hline
$569-G22\;\;\;\;$ & $Sc\;\;\;\;$ & $19.06\;\;\;\;$ & $10.788\;\;\;\;$ \\
\hline
$249-G16\;\;\;\;$ & $Sc\;\;\;\;$ & $18.76\;\;\;\;$ & $10.457\;\;\;\;$ \\
\hline
$90-G9\;\;\;\;$ & $SBc\;\;\;\;$ & $18.7\;\;\;\;$ & $10.531\;\;\;\;$ \\
\hline
$580-G29\;\;\;\;$ & $Sc\;\;\;\;$ & $18.64\;\;\;\;$ & $10.388\;\;\;\;$ \\
\hline
$34-G12\;\;\;\;$ & $Sc\;\;\;\;$ & $18.92\;\;\;\;$ & $10.851\;\;\;\;$ \\
\hline
\end{tabular}
\end{table}

As it is seen in Fig. 1, all galaxies have the same behavior for
small radii and then their velocities become constant and separate
from the linear regime in different values which we call it the
saturation velocity, $V_{sat}$. Tully-Fisher relation indicates that
$V_{sat}$ and the galaxies I-band luminosity, $L_I$, are related by
\begin{equation}
V\sim L_I^\alpha \;\;\;\;\;r\gg r_x,
\end{equation}
where $\alpha=0.2994$ for our data. Note that, the small difference
between this relation and the Tully-Fisher relation is that the
average velocty in constant regime is chosen instead of the maximum
velocity.

There is another scaling law beside relations (1) and (2). This
relation is a power law between the crossover radius and the I-band
luminosity. Since the crossover radius satisfies Eq. (1) and the
saturation velocity satisfies both Eqs. (1) and (2), so
\begin{equation}
r_x\sim L_I^z,
\end{equation}
where the exponent $z=\alpha/\beta$. Galaxies with larger luminosity
have larger crossover radius. The logarithmic diagram of crossover
radius as a function of galaxies I-band luminosity is plotted in
Fig. 2 and the exponent $z$ is obtained as $z=0.4173$ for our data
which is equal to the ratio $\alpha/\beta$.
\begin{figure}
\includegraphics[scale=0.5]{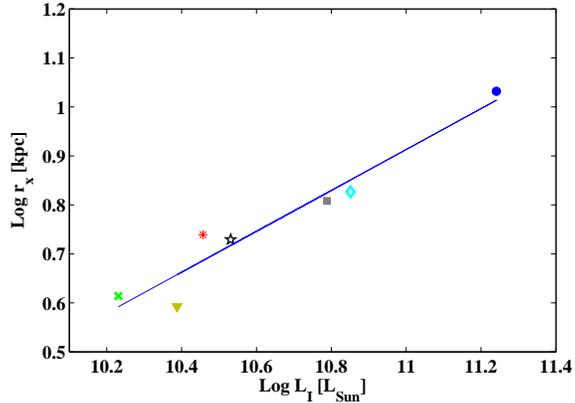}
\caption{Logarithmic diagram of crossover radius as a function of
galaxies I-band luminosity.}
\end{figure}

A universal scaling function can be constructed from our three
scaling relation as follow
\begin{equation}
V(r,L_I)=L_I^\alpha f(\frac{r}{L_I^z}),
\end{equation}
where the function $f(x)$ has the asymptotic behavior
\begin{eqnarray}
f(x)\sim\left\{\begin{array}{rcl}
&x^\beta\;\;\;\;\;\;\;\;x\rightarrow 0\\
&\rm const.\;\;\;\;\;x\rightarrow\infty.
\end{array} \right.
\end{eqnarray}
Actually, Eq. (4) governs the motion of stars in spiral galaxies.
This equation means that the data collapse onto a function $f(x)$
when plotted as $V/L_I^\alpha$ versus $r/L_I^z$ or correspondingly,
$V/V_{Sat}$ versus $r/r_x$ according to (2) and (3) (Fig. 3).
\begin{figure}
\includegraphics[scale=0.5]{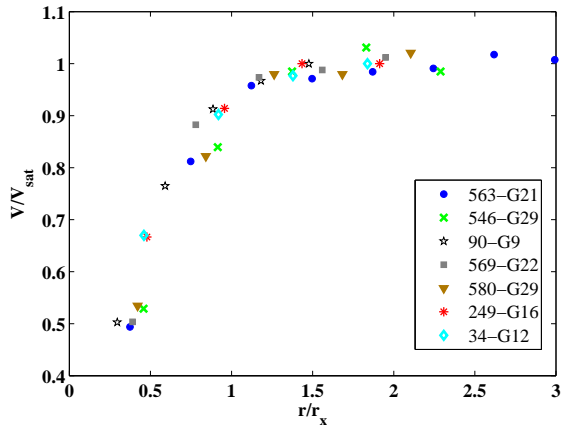}
\caption{Data collapse onto a universal function when plotted as
$V/V_{Sat}$ versus $r/r_x$.}
\end{figure}

Since, there is uncertainty in estimation of astrophysical
parameters and also existing empirical relations such as
Tully-Fisher relation have a considerable dispersion, it is
difficult to construct a tune universal function. Note that to study
these scaling relations together to find the universal function, we
need to pick galaxies whose RCs in logarithmic scale include both
linear and constant regimes.

This kind of diagrams and scaling functions exist in many areas of
physics such as surface growth \cite{Barabasi}, turbulent flow in
rough pipes \cite{Goldenfeld1}, non equilibrium phenomena and near
critical point problems \cite{Goldenfeld2, Kadanoff, Kardar}.
Generally, scaling functions are the signature of complex systems.

In addition to existing scaling function in galaxies there are other
evidences which lead us to consider galaxies as complex systems. For
instance there are mutual forces and interactions between stars in a
galaxy. Also, random velocities \cite{Schneider} and fluctuations
exist, especially in the core, and the number of stars in a galaxy
\cite{Schneider} is too much (ranged from $10^9$ to $10^{12}$).

Due to the collective behavior and interactions between the
components of a complex system, emergent and coherent structures
occur. Unlike the traditional scientific approach, the emergent and
coherent structures cannot be described within the vocabulary
applicable to the parts \cite{Goldenfeld3}. Dynamics of a system in
coherent state depends only on the nature of components interaction
and the equation of the motion of its components, lower level of
description, in a sense is violated. In fact, emergent levels of
description absorb the properties of lower level of description into
the phenomenological parameters and has own new law of physics. In a
paper entitled 'More is different' P.W. Anderson has described that
reductionism approach is not valid in emergent phenomena of complex
systems \cite{Anderson}. In this case the whole is more than the sum
of its parts. For further clarity, consider emergence of solid state
from the fluid state \cite{Goldenfeld4}. The pairwise interaction
between atoms does not change, but the correlation between them
changes when the temperature is lowered. Level of description of
solid state is described by continuum mechanic and has own new law
of physics. Origin of this new law is the collective behavior of the
atoms. Collective state exhibits novel response characteristics, and
loses memory of underlying level of description. It is why it took
so long for the existence of atoms to be deduced. The elastic forces
in solid are characterized by phenomenological parameters such as
shear modulus that can be computed from the atoms level of
description.

If we use star level of description for a spiral galaxy then its
arms would wind up soon as the galaxy rotates. Also, it is not
possible to describe the motion of stars in a spiral galaxy via a
few number of particles. Suppose we have a two body motion, one is
the star and the other is the total interior mass located at the
galaxy center. In this case we can not explain circular orbits of
stars in the disk sector according to this two body motion in which
the orbit should be a high eccentric elliptic. In fact the approach
to study complex systems is based on considering interaction and
collective effects. For instance, superconductivity cannot be
explained by just thinking about atoms. It is understood by
considering the cooperative behavior.

In galaxies moving stars with larger velocity try to accelerate
stars with smaller ones and vice versa. Thus interactions cause
irreversible transfer of momentum from points where the velocity is
large to the points where it is small and try to diminish the
velocity gradient. This correlation is the physical origin of
constant velocity in spiral galaxies.

In addition, stars in elliptical galaxies are on orbits that are
randomly oriented within the galaxy. On the other hand, because of
the existence of central mutual gravitational forces between stars
in a galaxy, they have a tendency to orbit in a unique plane
\cite{Thornton}. This kind of ordering and symmetry breaking
motivates us to consider that the disk galaxies emerge from
elliptical galaxies. In the other words, it seems that elliptical
and disk galaxies are two states of self gravitational systems. The
difference between them should be related to the correlation length
between their stars. Also, self gravity between stars in the disk
puts them in spiral arms when they rotate around the galaxy center
and spiral structure forms.

Thus, explanation of constant velocity in spiral galaxies needs
focusing on interaction between their components, stars, instead of
interpretation based on reductionism approach in lower level of
description which leads to exotic dark matter or modification of
gravitational force.

\textbf{Acknowledgement}

The author is grateful to Z. Rezaei, E. Kourkchi, M. Mehrafarin and
Sh. Rouhani for fruitful discussions.


\begin{thebibliography}{widest-label}
\bibitem{Salpeter} E.E. Salpeter, IAUS {\bf 77}, 23 (1978).
\bibitem{Sancisi} R. Sancisi, R.J. Allen, Astron. Astrophys. {\bf 74}, 73 (1979).
\bibitem{Krumm} N. Krumm, E.E. Salpeter, AJ {\bf 84}, 1138 (1979).
\bibitem{Moorsel} G.A. van Moorsel, Astron. Astrophys. {\bf 107}, 66 (1982).
\bibitem{Rohlfs} K. Rohlfs, R. Boehme, R.K. Chini, J.E. Wink, Astron. Astrophys. {\bf 158}, 181 (1986).
\bibitem{Zwicky} F. Zwicky, APJ {\bf 76}, 217 (1937).
\bibitem{Rubin} V.C. Rubin, W. K. Jr. Ford, and N. Thonnard, APJ {\bf 238}, 471 (1980).
\bibitem{Bosma} A. Bosma, AJ {\bf 86}, 1825 (1981).
\bibitem{Ashman} K.M. Ashman, PASP {\bf 104}, 1109 (1992).
\bibitem{Milgrom} M. Milgrom, APJ {\bf 270}, 365 (1983).
\bibitem{Sanders} R.H. Sanders and S.S.  McGaugh, Ann. Rev. Astron. Astrophys. {\bf 40} 263 (2002).
\bibitem{Brownstein} J.R. Brownstein and J.W. Moffat, APJ {\bf 636}, 721 (2006).
\bibitem{Moffat} J.W. Moffat, J. Cosmology Astropart. Phys., {\bf 5}, 3 (2005).
\bibitem{Tully} R.B. Tully, J. R. Fisher, Astron. Astrophys. {\bf 54}, 661 (1977).
\bibitem{Schneider} P. Schneider, {\it Extragalactic astronomy and cosmology}, Springer-Verlag (2006).
\bibitem{Courteau} S. Courteau, A.A. Dutton, F.C. Busch, L.A.M. Arthur, A. Dekel, D.H.M. Intush and D.A. Dale, APJ {\bf 671}, 203 (2007).
\bibitem{Shen} S. Shen et al. , ApJ {\bf 705}, 1496 (2009).
\bibitem{Persic} M. Persic, P. Salucci and F. Stel, Mon. Not. R. Astron. Soc. {\bf 281}, 27 (1996).
\bibitem{Barabasi} A.L. Barabasi and H.E. Stanly, {\it Fractal concept in surface growth}, Cambridge University Press (1995).
\bibitem{Goldenfeld1} N. Goldenfeld, Phys. Rev. Lett. {\bf 96}, 044503 (2006).
\bibitem{Goldenfeld2} N. Goldenfeld, {\it Lectures on phase transition and the renormalization group}, Perseus Books (1992).
\bibitem{Kadanoff} L.P. Kadanoff, {\it Statistical physics: Statics, dynamics and renormalization}, World Scientific (1999).
\bibitem{Kardar} M. Kardar, {\it Statistical physics of fields}, Cambridge University Press (2007).
\bibitem{Goldenfeld3} N. Goldenfeld and L.P. Kadanoff, Science {\bf 284}, 87 (1999).
\bibitem{Anderson} P.W. Anderson, Science {\bf 177}, 393 (1972).
\bibitem{Goldenfeld4} N. Goldenfeld, {\it Emergent state of matter}, Lecture notes (2011).
\bibitem{Thornton} S.T. Thornton and J.B. Marion, {\it Classical dynamics of particles and systems}, Thomson Learning, Inc. (2004).

\end{thebibliography}
\end{document}